\documentclass[useAMS,usenatbib]{mn2e}

\usepackage{graphicx}

\title[Berkeley94 and Berkeley96]{Berkeley 94 and Berkeley 96: Two Young Clusters with Different Dynamical Evolution
\thanks{Based on observations made with the Nordic Optical Telescope, operated on the island of La Palma jointly by Denmark, Finland, Iceland, Norway, and Sweden, in the Spanish Observatorio del Roque de los Muchachos of the Instituto de Astrof\'\i sica de Canarias.}}
\author[A.J. Delgado et al.]{A. J. Delgado$^{1}$\thanks{E-mail:
delgado@iaa.es}, A.A Djupvik$^{2}$, M.T. Costado$^{1}$ and E.J. Alfaro$^{1}$\\ $^{1}$Instituto de Astrof\'\i sica de Andaluc\'\i a~ (IAA-CSIC), Glorieta de la Astronom\'\i a, 18008-Granada, Spain.\\ $^{2}$Nordic Optical Telescope (NOT), Apdo. 474, 38700 Santa Cruz de La Palma, Spain.}

\begin{document}

\date{Accepted. Received; in original form}

\pagerange{\pageref{firstpage}--\pageref{lastpage}} \pubyear{2013}

\maketitle

\label{firstpage}

\begin{abstract}
We have performed multiband $UBVR_{C}I_{C}JHK_{S}$ photometry of two young clusters located at large Galactocentric distances in the direction of the Perseus spiral arm. The obtained distances and colour excesses amount to 3.9$\pm$0.11 kpc, $E(B-V)$=0.62$\pm$0.05 for Berkeley\, 94, and 4.3$\pm$0.15 kpc, $E(B-V)$=0.58$\pm$0.06 for Berkeley\, 96. The respective ages, as measured from the comparison of the upper colour-magnitude diagrams to model isochrones, amount to Log$_{10}$Age(yr)=7.5$\pm$0.07, and 7.0$\pm$0.07, respectively. A sequence of optical PMS members is proposed in both clusters. In addition, samples of objects showing $(H-K_{S})$ excess are found. Part of these are suggested to be PMS cluster members of lower mass than the optical candidates. The spatial distribution of these sources, the comparison to galactic models and to the expected number of contaminating distant red galaxies, and the spectral energy distribution in particular cases support this suggestion. The spatial distributions shown by members in different mass ranges can be interpreted in terms of the results from numerical simulations. According to these, different initial conditions and evolutionary dynamical paths are suggested for the clusters. Berkeley\,94 would have formed under supervirial conditions, and followed the so-called warm collapse model in its evolution, whereas Berkeley\,96 would have formed with a subvirial structure, and would have evolved following a cold collapse path. Both processes would be able to reproduce the suggested degree of mass segregation and their spatial distribution by mass range. Finally, the mass distributions of the clusters, from the most massive stars down to PMS stars around 1.3 M$_\odot$, are calculated. An acceptable general agreement with the Salpeter IMF slope is found.

\end{abstract}


\section{Introduction}

The observation of Young Open clusters in the Galaxy is the best tool to examine two issues of interest. First, whether there are differences in the star formation products observed at different locations in the Galactic disk. These differences are best checked in objects where the star formation is recent. Young clusters of ages around 10 Myr are best suited for the analysis, since star formation is no longer intensely active and  they are reliably observable at both optical and infrared wavelengths. On the other hand, they are still young enough that the observed properties are expected to fairly reflect the main properties of the star formation process, such as initial mass function, presence of massive stars, mass segregation, and spatial distribution of cluster members. Second, the star cluster observation is commonly accepted as the most adequate way to check models of stellar evolution. Young clusters in particular allow us to improve the evolutionary models of Pre-main sequence stars (PMS stars in the following).

In our project we concentrate on the observation with $UBVRIH\alpha+JHK$ photometry of clusters in the mentioned age range, with the aim of obtaining PMS member samples, mass distributions reasonably complete down to 1\,M$_\odot$, and an estimate of possible members for lower masses, (Delgado, Alfaro \& Yun 2011, and their references). This project connects with the aims of the currently ongoing GAIA-ESO Survey (Gilmore et al. 2012), where several of our southern objects of interest are included. Our observations furthermore should enable us to extend the analysis to objects whose membership estimates are expected to highly improve from GAIA data, but would not be accessible to spectroscopic observation with the present observational capabilities. 

\begin{figure}
   \centering
   \includegraphics[width=8cm,height=8cm,angle=0]{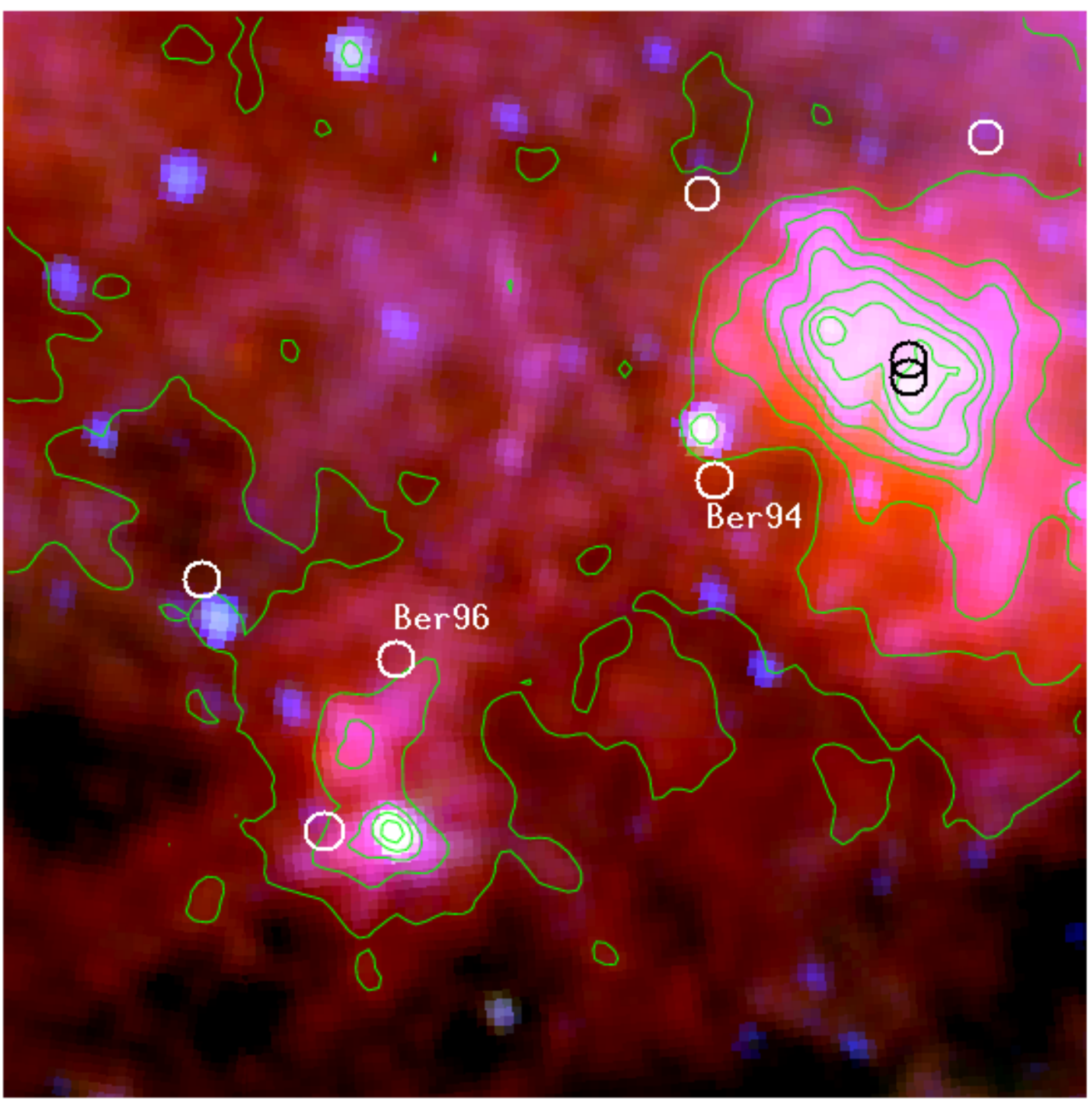}
      \caption{RGB map of 3 deg$^2$, combining the IRIS bands 60$\mu$\,(R), 25$\mu$\,(G) and 12$\mu$\,(B) from SkyView data base. Coordinates of the center are RA=22h\,25m\,53s, Dec=+55$^{\circ}$\,37'\,30''. North is up, east left. White and black circles indicate the positions of open clusters in the area, from Dias et al. (2002) catalogue, with labels for Ber94 and Ber96. Some IRAS point sources are visible, with the two brightest ones close to the locations of either cluster (see text). Contour levels (5 to 305 Mjy/sr) from the 60$\mu$ image are overplotted.}
         \label{mapa}
   \end{figure}

This refers in particular to the clusters Berkeley\,94 and Berkeley\,96, abbreviated in the following as Ber94 and Ber96. These objects are of particular interest in the context mentioned above, since they have been poorly observed up to now, and are located at relatively far galactocentric distances, in the direction towards the Perseus spiral arm of the Milky Way. Ber94 and Ber96 appear to be physically connected to a clusters concentration in the Perseus arm, whose more conspicuous object is the star forming and associated H{\sc ii} region SH\,2-132 (Saurin, Bica \& Bonatto 2010). Figure\,\ref{mapa} shows a field of 3 deg$^2$ with an R(60$\mu$)\,G(25$\mu$)\,B(12$\mu$) map from SkyView\footnote{http://skyview.gsfc.nasa.gov}, using IRIS bands. 

Ber96 (see Fig.\,\ref{mapa}), has a closest {\sc iras} source, {\sc iras} 22280+5515, apparently associated to a faint B0 star, TYC-3987-1285-1. No further information is contained in either catalogue about this source. 
Ber94 was included in the above cited study of the clusters related to the Sharpless region Sh\,2-132 (Saurin et al. 2010). Their study concentrates in several clusters actually embedded in the region, or located in its surroundings, and is based on 2MASS data for all them,  Ber94 included. In this context it would be part of the first generation of stars in a sequential formation process, whose last products are the young embedded clusters in the Sh\,2-132 region. On the other hand, Ber94 lies relatively close in projection to {\sc iras}\,22212+5542 (see Fig.\,\ref{mapa}), which is strongly dominated by the K-type red hypergiant and irregular variable RW\,Cep, HD\,212466. It is subject of further study whether some population of low mass stars is hidden in this region, and to what extent Ber94 would be physically associated with it.

\begin{figure}
   \centering
   \includegraphics[width=8cm,height=11cm,angle=0]{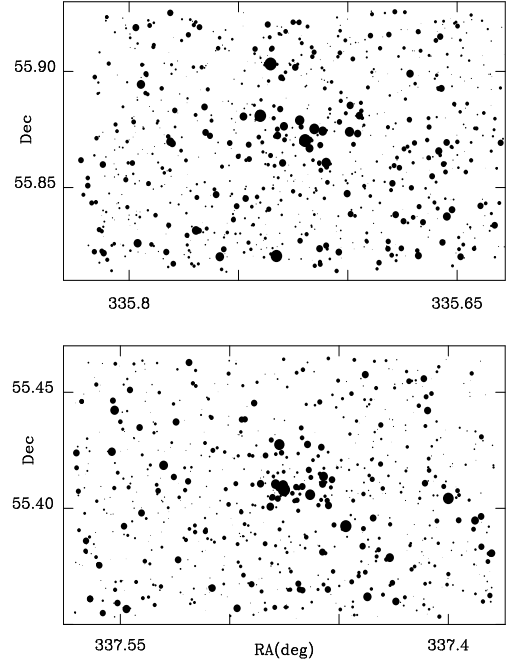}
      \caption{Schematic maps of the fields observed in Ber94 (upper panel) and Ber96. Right ascension and Declination are given in degrees. The dots sizes are coded according to the stars $V$ magnitude. Only stars with $V<20$ are plotted.}
         \label{mapas}
   \end{figure}

Equatorial (Epoch  2000) and Galactic coordinates are for Ber94: RA=22h\,22m\,42s, Dec=+55$^{\circ}$\,51'\,00'', l=103$^{\circ}$.095, b=-1$^{\circ}$.185; for Ber96: RA=22h\,29m\,24s, Dec=+55$^{\circ}$\,24'\,00'', l=103$^{\circ}$.661, b=-2$^{\circ}$.066. Their distance estimates put them well beyond the solar radius. Figure\,\ref{mapas} shows an schematic view of both fields, in which similar sizes for the clusters cores, of the order of 3' can be guessed. Their appearance is young, as they seem to be concentrated around a few bright stars that dominate the cluster brightness, and provide adequate targets to inquire into the star formation processes at large Galactocentric distances. In particular, we delve into the search and study of massive stars, PMS stars of intermediate and low mass, and relationship between these populations as members of the same cluster. In Section\,2, the observations and reduction are described. In Section\,3, the results are presented and analyzed. Section\,4 contains a summary of the main conclusions.

%


\section{Observations and reduction}

\subsection{Optical UBVR$_C$I$_C$H$\alpha$ observations}

The $UBVR_{C}I_{C}H_{\alpha}$ observations were obtained in available slots over several nights during September and December 2007, September 2010, July and September 2011. ALFOSC at the Nordic Optical Telescope (NOT) in the Roque de los Muchachos observatory (La Palma, Spain) was used. The detector was a 13.5\,$\mu$m$\times$2048$\times$2048 pixels, back-illuminated E2V CCD, covering a FOV of 6'.5 with a pixel scale of 0.19''. Long and short exposures were obtained in every band to cover as completely as possible the magnitude range of expected cluster members. {\sc iraf}\footnote{The Image Reduction and Analysis Facility (IRAF) is distributed by the national Optical Astronomical Observatory, which is operated by the Association of Universities for Research in Astronomy, Inc. (AURA) under cooperative agreement with the National Science Foundation.}  packages were used for image reduction. 

The calibration in the $UBV$ colors were performed through zero-point shifts, plus a significant colour term, calculated from the stars in common with published photoelectric photometry for both clusters, as given in WEBDA\footnote{http://www.univie.ac.at/webda//webda.html}  (Wramdemark 1976, 1978; del Rio 1984). For the calibration of Ber96 colours the stars 7 and 8 in WEBDA were excluded. They exhibit largely deviating values in the $V$ and $(U-B)$ indices, probably originating in variability or blending errors in the photoelectric values. Standards stars (Landolt 1992) in the fields SA\,107, SA\,114, PG\,2213, and PG\,2336 were observed  for calibration purposes in the $(V-R_{C})$ and $(V-I_{C})$ colour indices. The respective uncertainties of the calibrations of $V$, $(U-B)$, $(B-V)$, amount to 0.04, 0.02, 0.02\,mag, in Ber94 and 0.04, 0.05, 0.04\,mag in Ber96. In the colour indices $(V-R_{C})$, and $(V-I_{C})$ the calibration errors reach up to 0.04\,mag and 0.05\,mag, respectively.

The four brightest stars in the field of Ber94 deserve further comments. Star 3930 (WEBDA\,1) is a foreground GIII giant, as listed in the Tycho catalogue (H\o g et al. 2000) and is excluded as a cluster member. The other three stars brighter than $V\simeq$11\,mag are saturated even in the shortest $B$-band exposure. Star 739 is not included in WEBDA. It is catalogued as a B0 star by Reed (2003) and has Tycho identity TYC 3986-2178-1. Stars 3903 and 3904, (WEBDA 23 and 38 respectively) have WEBDA $UBV$ photometry from Wramdemark (1978), and $B$ values from the Tycho catalogue. The available WEBDA indices, and the location of these stars in the three available colour-magnitude (CM) diagrams, $V$ vs $(U-V)$, $(V-R_{C})$ and $(V-I_{C})$, suggests their cluster membership. 

\subsection{Near Infrared JHK$_{S}$ observations}

The $JHK_S$ images were obtained with NOTCam\footnote{The Nordic Optical Telescope's Infrared Camera and Spectrograph, for more details see http://www.not.iac.es/instruments/notcam/ .} at the NOT, La Palma, using the wide field camera (0.234"/pixel, field of view 4' $\times$ 4') on 2 September 2007. The detector was the engineering grade Hawaii 18.5 $\mu$m$\times$1024$\times$1024 pixels HgCdTe array (SWIR1). To avoid saturating on bright stars, Ber94 was imaged with an exposure time of 5 seconds, combining 72 dithered frames to 360 seconds of total integration time per filter, the seeing being of FWHM = 0.9". For Ber96 we used 7.2 seconds exposure time and combined 36 dithered frames to get 260 seconds per filter in slightly better seeing conditions, FWHM=0.8". Even so, the three brightest stars in each cluster exceed the linear range of this detector and their NOTCam flux is excluded.

The data was reduced using {\sc iraf} with a set of own scripts optimized for NOTCam data. The raw data was corrected for bad pixels and divided by the master flats obtained from differential twilight flats. Then the sky was estimated from the dithered images and a properly scaled sky was subtracted from each individual frame. A distortion model for NOTCam was applied to correct the individual images for optical distortion before aligning and shifting. Aperture photometry with a radius of 4 pixels was made on all detected sources, and aperture corrections derived from a set of isolated, bright stars was applied. The 2MASS catalogue (Skrutskie et al. 2006) was used to calibrate photometry and astrometry. Positions are accurate to 0.08" in right ascension and declination for Ber94 and 0.11" for Ber96. We found a systematic shift in offset magnitudes with position on the array. In the lack of a proper illumination correction model, the photometry was calibrated with the empirical slope  (2$\times$10$^{-4}$\,mag/pixel in Y-direction) plus a median offset found from comparing with 2MASS sources of quality flag AAA. A total of about 30-40 calibration stars were used after dismissing doubles unresolved by 2MASS. The scatter between NOTCam and 2MASS magnitudes after calibration is 0.06\,mag in $H$ and 0.07\,mag in $J$ and  $K_{S}$. These are used as an upper conservative estimate of the uncertainty in the photometric calibration. The photometry of Ber96/Ber94 reaches a depth of 19.3/19.6, 18.9/19.0 and 18.0/18.3 magnitudes in $J$, $H$, and $K_S$, respectively, with an instrumental S/N ratio of 10. This is about 4 magnitudes deeper than 2MASS. For Ber94, a recent study contains an analysis of 2MASS data in its field (Saurin et al. 2010).

\section{Results and discussion}

\subsection{Membership of MS and evolved stars}

The membership of stars located in the upper part of the CM diagram is assigned through fitting the diagrams to the zero-age main sequence (ZAMS). In the calculations we assume that the interstellar reddening in the direction of the clusters is described by an average extinction law (Cardelli, Clayton \& Mathis 1989) with an absorption coefficient $R_{V}=A_{V}/E(B-V)=3.1$, and a reddening slope $E(U-B)/E(B-V)=0.72$. The procedure has been explained in detail by Delgado et al. (1998), and applied systematically to several clusters (Delgado, Alfaro \& Yun 2007, 2011, referred to in the following as DAYI and DAYII, respectively). This same procedure provides membership estimates for the most luminous stars in the CM diagram, and select those most adequate to measure the cluster age with respect to postMS isochrones.

\begin{figure}
   \centering
   \includegraphics[width=9cm,height=10.5cm,angle=0]{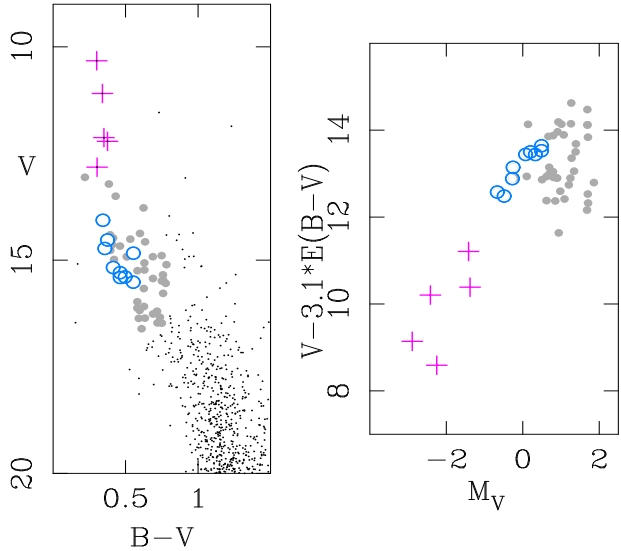}
      \caption{Selection of MS and postMS members in Ber96. The left panel shows the upper $V$,$(B-V)$ CM diagram, with the stars selected as possible MS members marked as dots, circles and crosses. In the right panel the quantity $V$-3.1$\times\ E(B-V)$ is plotted versus $M_{V}$ for these stars. The stars plotted as blue circles are selected as unevolved MS members. Magenta crosses are the stars selected as postMS members, used to calculate the cluster age.}
         \label{Acto96}
   \end{figure}

Figure\,\ref{Acto96} shows as example the procedure for main sequence (MS) and post-main sequence (postMS) members selection applied to Ber96. Stars plotted as grey dots and blue circles in both panels show the first broad selection of MS members, first performed in the $V$,$(B-V)$ CM diagram, and checked afterwards in the remaining CM and colour-colour (CC) diagrams to avoid clearly deviating stars in any colour index. The blue circles and magenta crosses in Fig.\,\ref{Acto96} are, respectively, the selected unevolved MS cluster members, and the postMS members used to calculate cluster age, according to their location in these plots, and also in the other CM diagrams. The average values of colour excess and distance modulus for the unevolved MS members are adopted for the cluster. They amount to $DM$=12.97$\pm$0.14, $E(B-V)$=0.62$\pm$0.05 for Ber94 and $DM$=13.16$\pm$0.14, $E(B-V)$=0.58$\pm$0.06 for Ber96. These are used as reference values to estimate membership for all the stars.

\subsection{Membership of PMS stars}

The procedure to assign membership to PMS candidates is the same as for MS stars, but using PMS isochrones of ages 1 to 10 Myr, instead of the ZAMS, to measure distances of the candidate members. Here we use models by Siess, Dufour \& Forestini (2000), Tognelli, Prada-Moroni \& Degl'Innocenti (2011) and Yi et al. (2001). In the following we refer to them as S00, T11, and Y01, respectively. Membership is assigned when one or more of these calculated distances coincide with the average distance estimated for MS members. The $(U-B)$, $(B-V)$ CC  relation is the same as the one for ZAMS, which is a valid assumption for PMS stars of Class III and older (Delgado et al. 1998). The median color excess value of all the stars with valid $U$-band detection and membership assignment is used as reddening value to calculate the distance modulus of stars without valid $U$ measurement. The membership is checked in all four CM diagrams, available with our $UBVR_{C}I_{C}$ photometry. The final candidate PMS members are those stars with assignment in at least 3 CM diagrams, and with photometric error in $V$ and in all colour indices smaller than 0.05 mag. A detailed description of the procedure is given in DAYI, and exemplified in Figs.\,4 and 5 of this paper. 

As explained above, the standard calibration in the $UBV$ colour indices is based on zero point shift plus colour term from the comparison to published photoelectric photometry for the brightest stars in the field. The extrapolation to higher magnitudes introduces a higher uncertainty of the member selection for fainter selected members. This would mainly affect the lower part of our calculated mass distributions. To minimize this uncertainty, we may consider a reduced PMS members sample, which includes only those members with the same age from every CM diagram with assignment, inside 1 Myr difference. We recall that ten PMS isochrones are used, with 1\,Myr spacing between them. We refer to this reduced members sample as PMSR members. The procedure of PMS membership assignment provides ages for the candidate members, and an indication of age dispersion, from the median and mean deviation of all individual age values. These median and deviation amount to 7.0$\pm$2.4\,Myr and 6.5$\pm$2.2\,Myr for PMSR candidates in Ber94 and Ber96, respectively.

Two more comments are needed on the PMS members selection. Several stars, located at the position of the CM diagram where PMS and MS sequences should merge, have membership assignments both as MS members and as PMS members around the highest PMS ages considered, 8 to 10\,Myr. We keep as PMS candidates those among them fainter than an A0V star at the calculated distance and reddening, whereas those brighter than this limit are adopted as MS candidate members. On the other hand, some bright stars are assigned as highly reddened PMS members in the youngest isochrones considered, around 3 to 1\,Myr. In the CM diagrams they are located at similar $V$ as  the MS candidates. Some unquantifiable contamination might be present here, by binary stars, highly reddened MS members, or even foreground stars of later spectral type. 

We finally mention here the absence of any $(R-H\alpha)$ excess in the proposed PMS sequences of both clusters. A single exception is star 83 in Ber96, which we comment later, in Sect.\,3.5 below. The index $(R-H\alpha)$ can be used as indicator of $H\alpha$ emission, thus helping in the identification of possible PMS cluster members (Sung, Chun \& Bessell 2000). In the present case, no sign of an excess in this index is found. This indicates the practical absence of any remainings of disk or envelope material around the PMS candidates, as compared to the results found in younger clusters (Yun et al. 2008). This absence, together with the lack of other possible PMS features, such as $(H-K_{S})$ near infrared (NIR) excess, strongly depends on the age of the PMS candidates, and is actually expected in clusters of ages similar to those obtained for Ber94 and Ber96.

\subsection{Optical Photometric diagrams}

The optical catalogue is matched with 2MASS data, to obtain astrometry with precission of 0.08 arcsec in both coordinates. Together with the matching of the NIR photometric catalogue (see Sect.2.2 above) a complete catalogue of $UBVR_{C}I_{C}$ and $JHK_{S}$ colours is obtained. It will available as a catalogue at CDS. 

\begin{figure}
   \centering
   \includegraphics[width=10cm,height=12cm,angle=0]{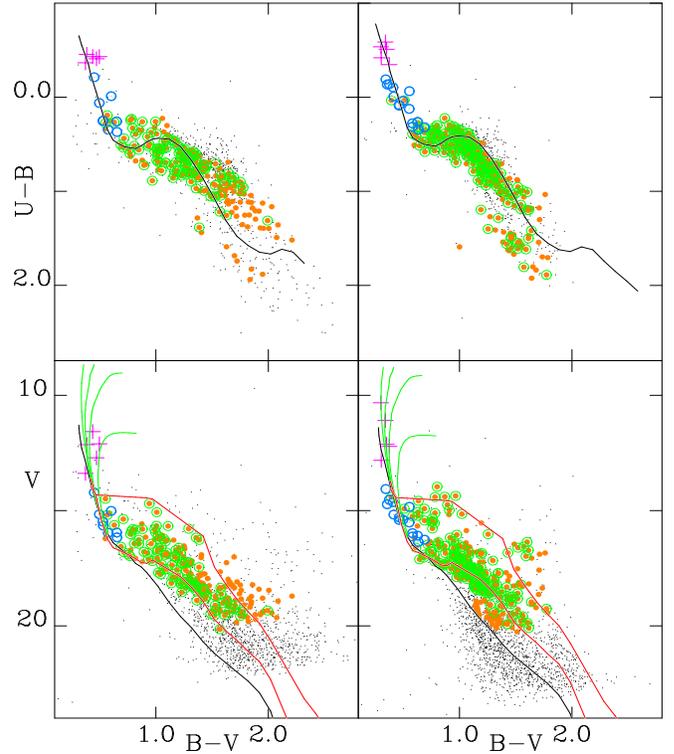}
      \caption{CC an CM diagrams for Ber94 (left panels) and Ber96. $(U-B)$,$(B-V)$ (upper panels) and $V$,$(B-V)$ diagrams are shown. The ZAMS (black line), postMS isochrones for LogAge(yr) 6.6, 7.0, 7.4 and 8.0 (green lines), and PMS isochrones for 10\,Myr and 1\,Myr (red lines) are plotted to account for the clusters colour excesses and distances. Evolved members used in the age calculation (magenta crosses), MS members (blue circles) and assigned PMS members (orange dots) are indicated. Green circles mark indicate the so-called PMSR members (see text, Sect. 3.2).}
         \label{CCCM-3}
   \end{figure}

In Figure\,\ref{CCCM-3}, we show in various panels the $(U-B)$,$(B-V)$ CC diagram and the $V$,$(B-V)$ CM diagrams for both clusters. The postMS evolved members, used to calculate the cluster age, MS members, and PMS members are indicated. In all diagrams the ZAMS line (Schmidt-Kaler 1982), postMS isochrones (Girardi et al. 2002; G02 in the following) and PMS isocrones (S00) are shifted to account for the calculated colour excesses and distance moduli.  

The cluster age is calculated from comparison of the selected bright stars to the G02 isochrones in all CM diagrams. For Ber94, where the three brightest stars do not have $B$-band measurement, we use the $V$,$(U-V)$ CM diagram. The estimated age for each star is interpolated in every CM diagram, and the adopted value is the average of all values. The uncertainty is calculated as the standard deviation of this average, divided by the square root of the number of stars used in the estimate (see DAYII for details). The resulting values amount to Log$_{10}$Age(yr)=7.5$\pm$0.07, and 7.0$\pm$0.07, for Ber94 and Ber96 respectively.

\subsection{NIR photometric diagrams. Infrared excess}

\begin{figure}
   \centering
   \includegraphics[width=9cm,height=11cm,angle=0]{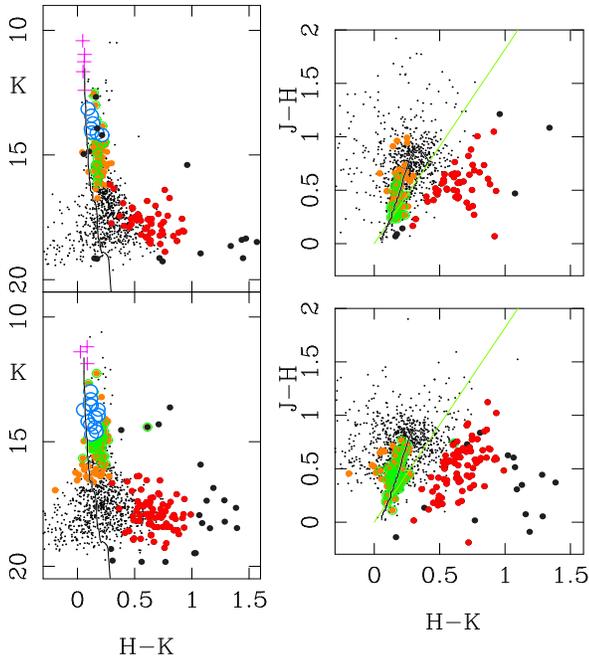}
      \caption{NIR photometric diagrams, $K_{S}$,$(H-K_{S})$ and $(J-H)$,$(H-K_{S})$ for Ber94 (upper panels row) and Ber96. The ZAMS line (black line) is plotted to account for the reddening and distance to the clusters. The green straight lines mark the reddening direction in the CC diagrams. Symbols are as in Fig.\ref{CCCM-3}. Red dots and larger black dots are sources with NIR $(H-K_{S})$ excess (see text), as selected in the $(J-H)$,$(H-K_{S})$ CC diagrams. Black dots are NIR-excess sources rejected to calculate maps of spatial distribution, chosen from their location in the $K_{S}$,$(H-K_{S})$ CM diagrams.}
         \label{NIRCCCM}
   \end{figure}

In Figure\,\ref{NIRCCCM} we show the $(J-H)$,$(H-K_{S})$ CC diagrams and $K_{S}$,$(H-K_{S})$ CM diagrams for both clusters.  Lines of reddening for the NIR extinction law by Whittet (1988) adapted to 2MASS bands are plotted. The slope is $E(J-H)/E(H-K_{S})$=1.8. We also mark the sources with NIR excess, as those with separation from the reddening line larger than their uncertainty in both colour indices plotted. These uncertainties are calculated as squared sums of the photometric errors plus the errors of the zero point calibrations with respect to 2MASS sources (see Sect. 2.2). The NIR-excess sources are also marked in the $K_{S}$,$(H-K_{S})$ CM diagrams, compared to the position of the optically selected PMS candidate members. 

\begin{figure}
   \centering
   \includegraphics[width=9cm,height=11cm,angle=0]{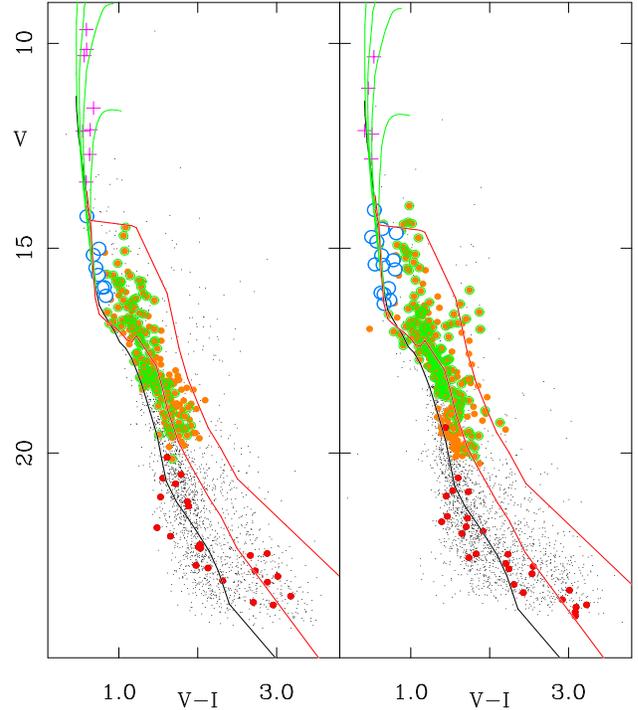}
      \caption{$V$,$(V-I_{C})$ CM diagram for Ber94 (left panel) and Ber96. Symbols for the different candidate members, and ZAMS, postMS and PMS isochrones, are plotted as in Fig.\,\ref{CCCM-3}. The red dots in the lower part of the diagram represent those NIR-excess sources shown in Fig.\,\ref{NIRCCCM}, which are also detected in optical wavelengths.}
         \label{V-VI}
   \end{figure}

As it is apparent from the plots in Fig.\,\ref{NIRCCCM} almost no stars are at the same time optical PMS candidate members and NIR-excess sources. Among these last sources, very few have an optical detection, and are at the same time located in favourable positions in the optical CM diagrams, at $V$ values fainter than $V\approx 20$\,mag. In Figure\,\ref{V-VI} we show the $V$,$(V-I_{C})$ CM diagram, with indication of the assigned members. The lines and symbols are as in Fig.\,\ref{CCCM-3}. Fig.\,\ref{V-VI} includes as red dots those stars among the above selected NIR-excess sources, which also have optical colours. Their location in this optical CM diagram suggests that some of them could be low mass cluster members. This is discussed below.

\subsection{Cluster membership of NIR-excess sources}

NIR excess calculated from the $(J-H)$,$(H-K)$ CC diagram is typically attributed to thermally radiating hot circumstellar dust, i.e. young T Tauri type stars with disks or even protostars with envelopes. These same NIR colours may, however, be displayed by distant, redshifted galaxies (Hewett et al. 2006). In order to assess the possible contamination by extragalactic sources we first compare the $K$-band histograms. 

\begin{figure}
   \centering
   \includegraphics[width=9cm,angle=0]{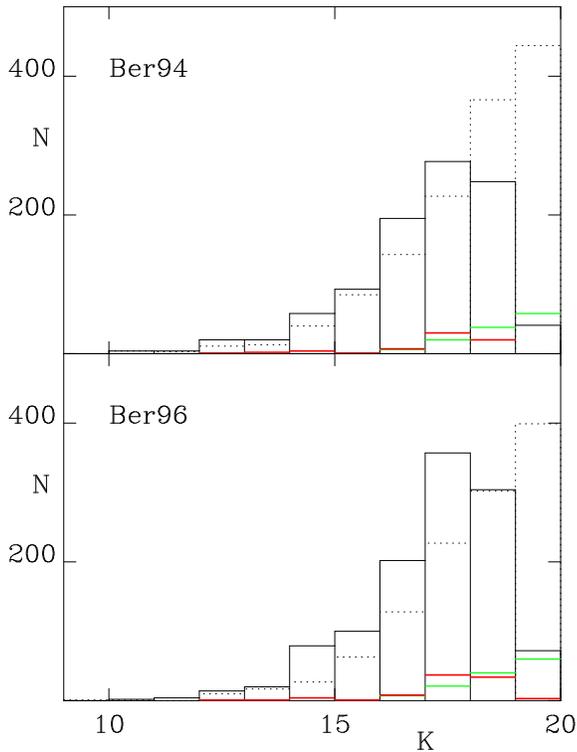}
      \caption{Differential $K$-band histogram for Ber94 (upper) and Ber96 (lower) showing the number of sources per $K$-band magnitude bin. As expected, the observed $K_{S}$-band counts (black continuous bars) are in excess of the counts from the Besan\c{c}on Galaxy model (black dotted bars) until approximately $K$=18\,mag, where incompleteness of the data sets in. The histograms of the NIR-excess sources (red bars) peak in the range 17 $< K <$ 19 magnitudes. Extragalactic $K$-band counts from the literature (green bars) are plotted for comparison. See text for details.}
         \label{khist}
   \end{figure}

Figure\,\ref{khist} shows differential $K$-band histograms for both clusters. In the same histogram we plot the total number of observed $K_{S}$-band sources (black continuous bars), the number of stars expected from the Besan\c{c}on model of our Galaxy (black dotted bars; Robin et al. 2003), the small subset of NIR-excess sources (red continuous bars), as well as the number of extragalactic $K$-band counts expected (green continuous bars). For the latter we used the average values of the deep fields counts in Table 4 of Barro et al. (2009) and scaled the numbers down to the size of our fields. In the bin centred on $K$=17.5\,mag the NIR-excess population is larger than what can be expected from extragalactic source counts for both clusters. In the bin centred on $K$=18.5\,mag, however, extragalactic sources can account for 15 - 20\% of all $K_{S}$-band counts, which is more than all the NIR-excess sources. It is clear otherwise, that there are more faint sources, and also more NIR-excess sources in the Ber96 field than in the Ber94 field. On the other hand, the histogram of absolute $J$-band magnitudes peak in both cases at $J$=5.5\,mag (corresponding roughly to the completeness limit), which indicates masses in the range 0.2-0.5\,M$_{\odot}$ according to the models of Baraffe et al. (1998), when assuming an age for the NIR-excess sources of 6\,Myr, similar to the optically selected PMS candidates.

The release of the $WISE$\footnote{Wide-field Infrared Survey Explorer, http://wise.ssl.berkeley.edu} All-Sky Source Catalog (Wright et al. 2010) allows us to extend the SED of some of our targets into the mid-infrared (mid-IR). We have cross-correlated our source list with the WISE catalogue, finding a total of 191 and 216 $WISE$ counterparts in Ber94 and Ber96, respectively. Among these, only 19 and 39 stars are optically selected PMS stars in Ber94 and Ber96, respectively, and none of these have mid-IR excess. Some cases of interest are discussed below.

   \begin{figure}
   \centering
   \includegraphics[width=9cm,angle=0]{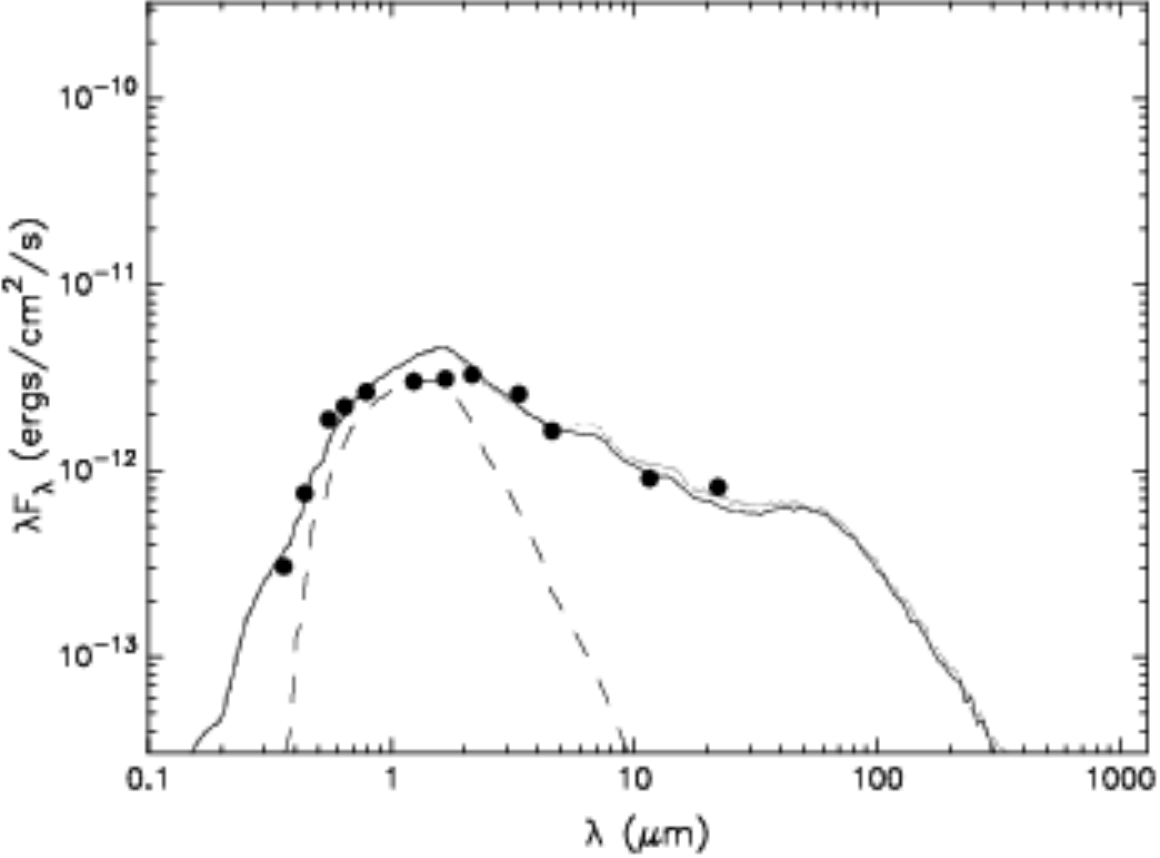}
      \caption{The SED sampling $UBVR_{C}I_{C}JHK_{S}$ and the four WISE bands of the source Ber96-474 is fitted to YSO models using the Online SED Fitter (Robitaille et al. 2007) and restricting the distance and the cloud extinction to that of Ber96. The best fit is that of a 1 Myr old 1.7 M$_\odot$ star. The dashed line represents the underlying stellar photosphere}
         \label{sed}
   \end{figure}

\subsubsection{Notes on particular sources}

From the results of SED fitting to NIR and mid-IR data, using the online SED fitting tool (Robitaille et al. 2007), we remark the cases of two stars of particular interest. The star Ber96-83 is the only optical PMS candidate with a NIR excess, and it also has an excess in the $(R - H_{\alpha})$ index. Our results from PMS membership assignment in Sect. 3.2 suggest a 1.6 M$_\odot$ star with an age of 9.5 Myr with a cloud extinction of $A_V$ = 1.8 mag. The YSO model that best fits the observed SED suggests a 1.7 M$_\odot$ star with an age of 7.7 Myr when the distance range is limited to 4.0-4.3 kpc and the visual extinction is in the range 1.7-1.9 mag.

On the other hand, only 4 and 2 stars, respectively, among the NIR-excess sources in Ber94 and Ber96 are detected with $WISE$, and star Ber96-83, commented above is not among them, but the particular case of a brighter star, Ber96-474, deserves a comment. It is the only star with valid measurement in all $UBVR_{C}I_{C}JHK_{S}$ colours, having NIR excess and also a clear mid-IR excess defined with the $WISE$ colour indices [3.4]\,-\,[4.6]\,$>$\,0.4\,mag and [4.6]\,-\,[12]\,$>$\,2.0\,mag. This infrared excess strongly suggests that the object is a YSO, which would mean that star formation is active in Ber96. This object indeed has an optical PMS membership assignment, but only in the $V$,$(B-V)$ diagram, and the lack of confirmation in the other diagrams caused our procedure to exclude it as a bona-fide PMS member. It is quite likely, though, that an IR-excess YSO produces scattering in the optical that would give abnormal broad-band colours. Restricting the distance and cloud extinction to that of Ber96, the best fit to the observed SED sampled from 0.37 to 22 $\mu$m (see Fig.\,\ref{sed}), is given by a 1.7 M$_\odot$ star with an age of about 1 Myr. Among the 10 best fits, 6 are with this model at various inclination angles, the remaining 4 being different models, all at inclination angles $>$ 80 degrees, masses from 2-5 M$_\odot$ and ages from 0.4 to 5.5 Myr. The observations can not be fit by a reddened photosphere.

\subsection{Spatial distributions}

The knowledge of the dynamical state of a stellar system requires the determination of both its mass and kinematical distributions. Proper motions and radial velocities are required for the adequate determination of the latter, but a partial approach can be obtained from several descriptors of the physical structure which relate the results of dynamical models to observations. In particular, descriptions of the spatial distribution of cluster members with the so-called parameter Q (Cartwright \& Whitworth 2004; S\'anchez \& Alfaro 2009), the surface stellar density, $\Sigma$ (see Bressert et al. 2010 and references therein), or some other parameter related to mass segregation (Allison et al. 2009; Olczak, Spurzem \& Henning 2011) provide information about the star formation history and cluster dynamics of the region (e.g Parker, Maschberger \& Alves de Oliveira 2012; Parker \& Meyer 2012).

\begin{figure*}
   \centering
   \includegraphics[width=4.5cm,height=4.5cm,angle=0]{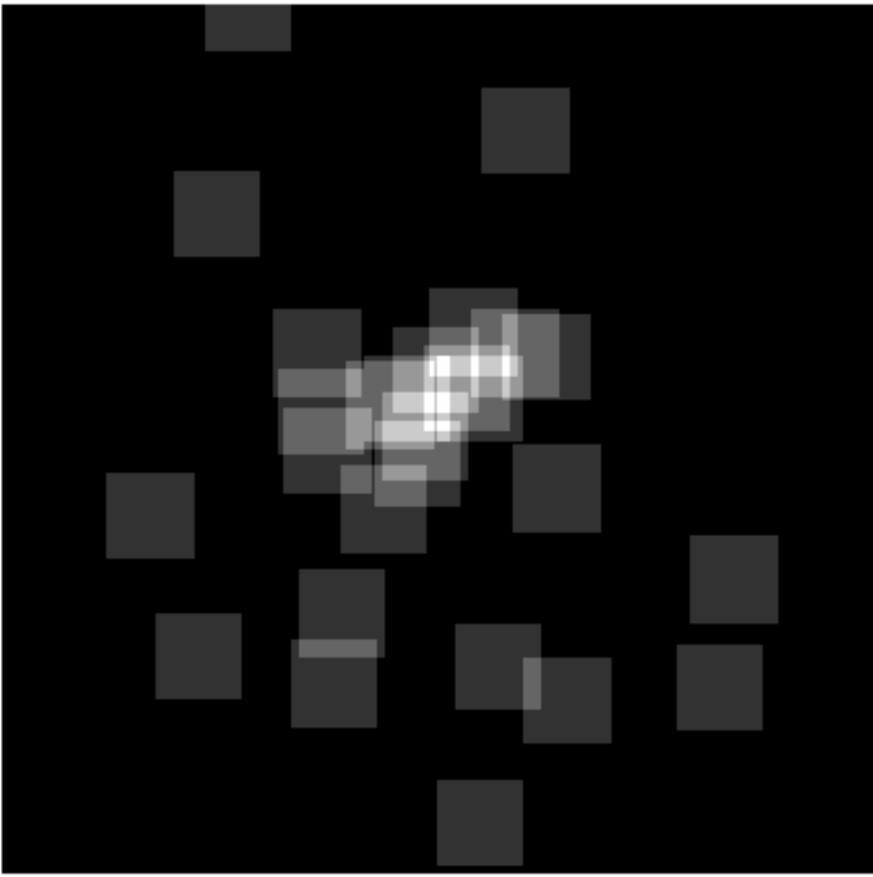}
   \includegraphics[width=4.5cm,height=4.5cm,angle=0]{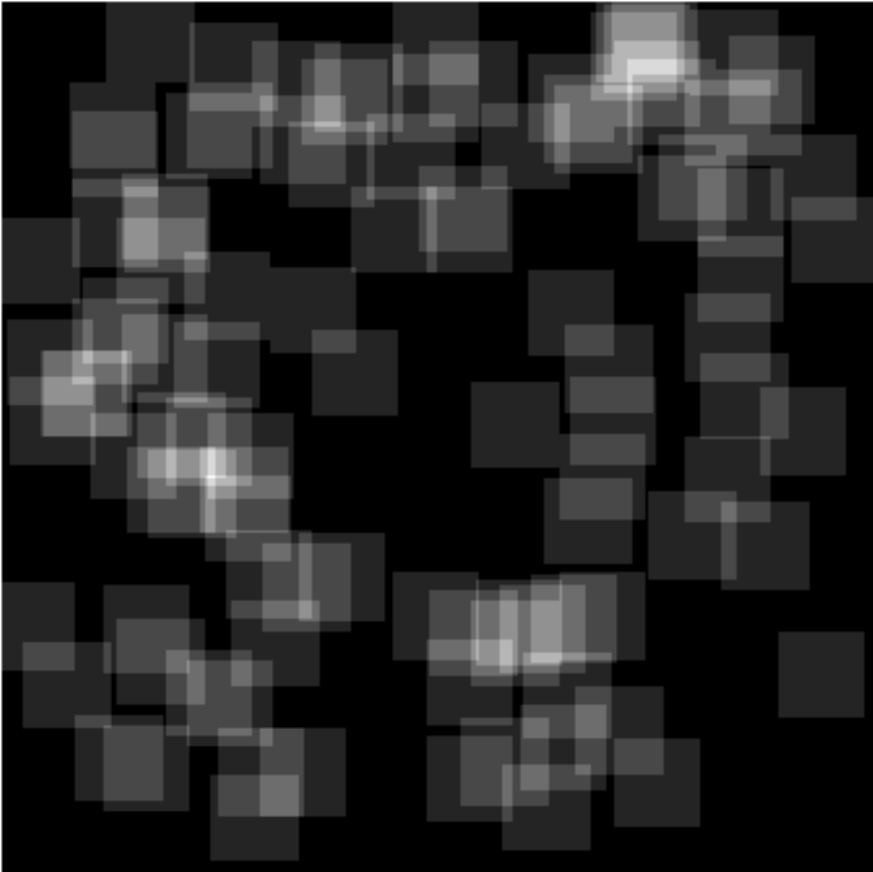}
   \includegraphics[width=4.5cm,height=4.5cm,angle=0]{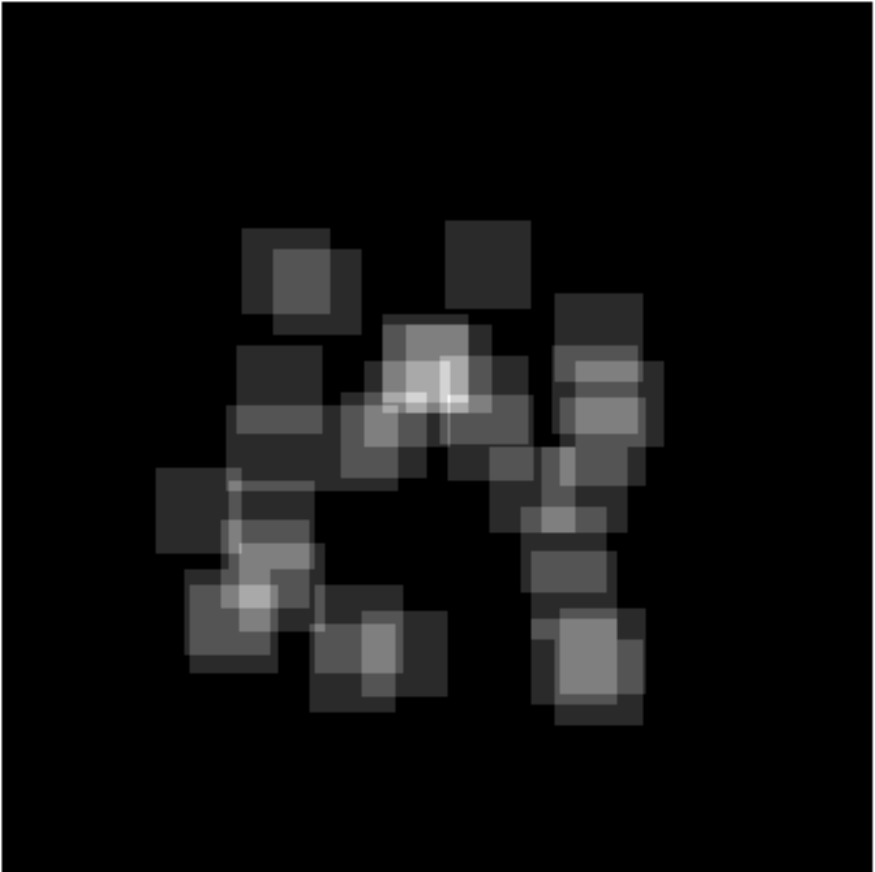}
   \includegraphics[width=4.5cm,height=4.5cm,angle=0]{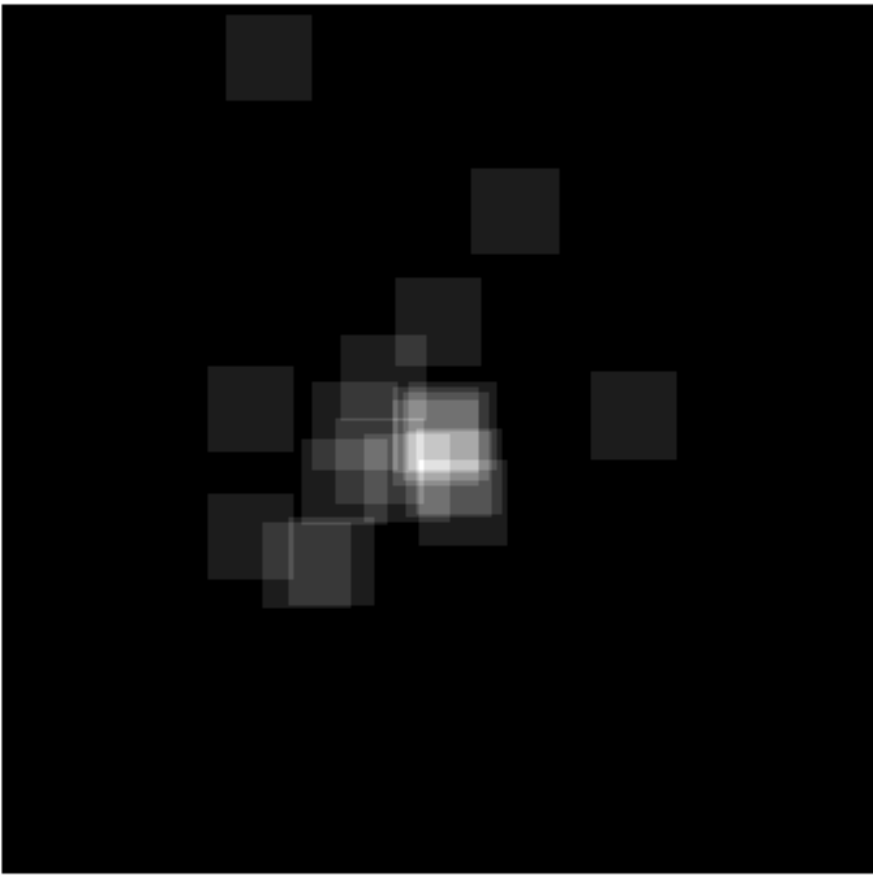}
   \includegraphics[width=4.5cm,height=4.5cm,angle=0]{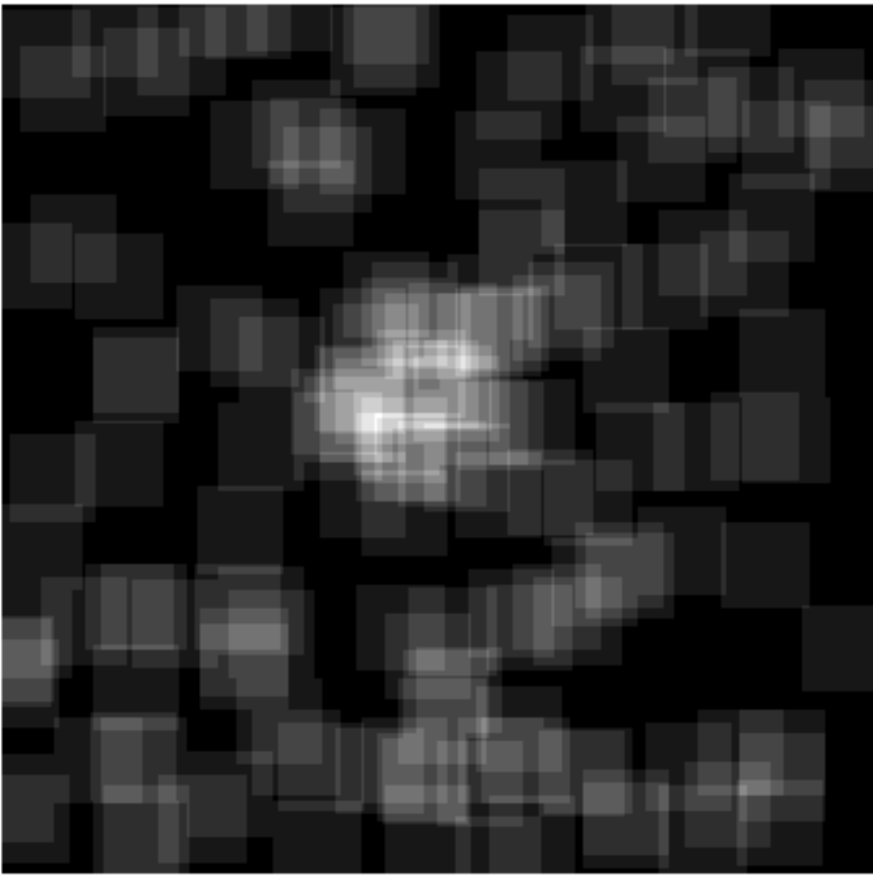}
   \includegraphics[width=4.5cm,height=4.5cm,angle=0]{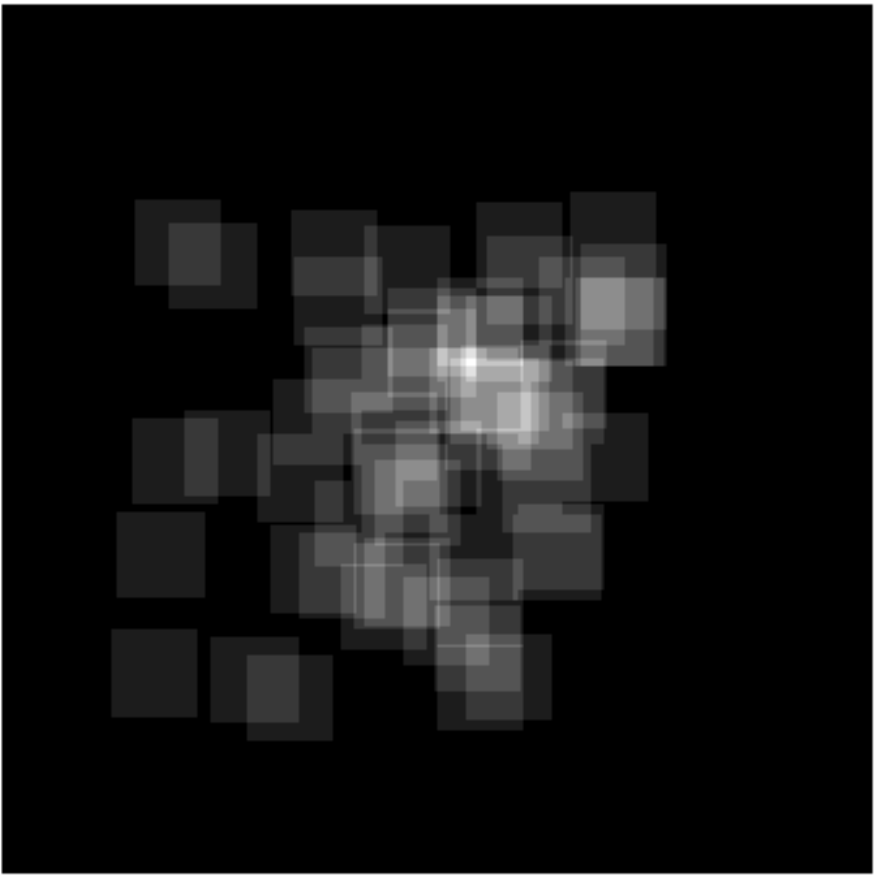}
      \caption{Density maps for different types of members. From left to right: MS+postMS candidate members, PMSR members, and NIR-excess sources. Upper row shows the maps for Ber94, bottom row for Ber96. North is up and East left. The plots show signs of comparatively tighter concentration around the cluster center in Ber96, particularly obervable in MS+postMS and PMS members. In both clusters, mainly in Ber94, signs of segregation of low mass members are observable. The field in the NIR maps (right panels) covers the central 4'$\times$4' in the optical field of view.  
              }
         \label{Spatial}
   \end{figure*}

Figure\,\ref{Spatial} shows from left to right the density maps for MS and postMS member stars, candidate PMSR members (Sect.\,3.2), and NIR-excess sources as shown in Fig.\,\ref{NIRCCCM}. The maps represent numbers of stars of each type inside 20$\times$20 arcseconds bins, normalized to the number in the maximum bin. Two features deserve attention. The selected MS and postMS members in Ber96 (lower left panel) show a clear central concentration. In addition, the selected PMSR members (lower middle panels) are also centrally condensed around the same location. This supports the reliability of the method of membership assignment. Even the spatial distribution of the NIR-excess sources appears concentrated in this location, which suggests that an important fraction of these sources could be  low mass cluster members, highly reddened and not visible at optical wavelengths. On the other hand, the maps in Fig.\,\ref{Spatial} show different distributions for candidate members in different mass ranges. Average masses amount to 7.92 $M_\odot$, 2.30 $M_\odot$, for the MS+postMS, and PMS members in Ber94, and 7.46 $M_\odot$ and 2.05 $M_\odot$ for the same members in Ber96. Average masses of the NIR excess sources plotted in the respective rightmost plots in Fig.\,\ref{NIRCCCM} are not quantitativeley determined. As commented above, those cluster members among these sources would have masses around 0.5 $M_\odot$.

\subsection{Mass segregation}

\begin{figure}
   \centering
   \includegraphics[width=10cm,height=12cm,angle=0]{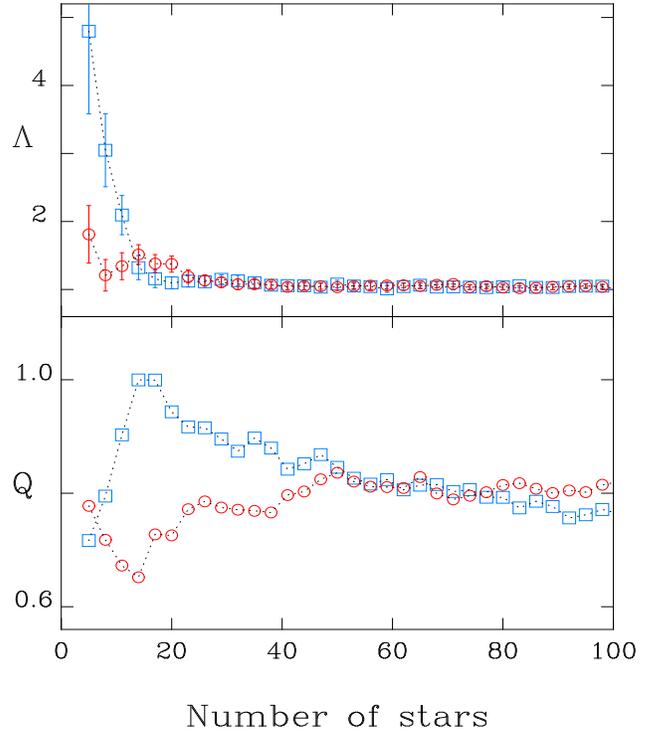}
      \caption{Variation of parameters $\Lambda$ and Q in both clusters, as function of the number of stars, counted in order of decreasing mass. Blue squares and red circles represent values for Ber96 and Ber94, respectively. $\Lambda$ values above unity indicate tighter concentration in the corresponding mass range. This is observed for the most massive stars in both clusters, and relatively stronger in Ber96. The variation of Q suggests different types of formation process and dynamical evolution for both clusters (see text). }
         \label{lambdaQ}
   \end{figure}

The parameter $\Lambda$, based on the so-called minimum spanning tree (MST), allows us to quantify the degree of mass segregation in a cluster and to examine how mass segregation changes for different ranges of stellar masses (Allison et al. 2009). The details of the spatial structure are described with the parameter Q (Cartwright \& Whitworth 2004) for several subsets of cluster members. Values of Q and $\Lambda$ have been calculated for different mass intervals, starting with the 5 most massive stars, and adding three members in order of decreasing mass in each new calculation. The results are plotted in Fig.\,\ref{lambdaQ}. The mass values in this calculation are from G02 models for MS and postMS stars, and S00 models for PMS stars. All MS, postMS and PMS candidate members are considered.

The upper panel in Fig.\,\ref{lambdaQ} represents the variation of $\Lambda$ with increasing star numbers. $\Lambda$ values close to one indicate absence of mass segregation in either sense. Values of this parameter above unity indicate the presence of a relative tighter concentration in the corresponding mass range. The values plotted in Fig.\,\ref{lambdaQ} suggest the presence of mass segregation in both clusters, in the sense of more concentrated distribution for the higher mass stars. Only the 17 most masive stars in Ber94 (24.2\,-\,4.6\,M$_{\odot}$) are contained inside the 70\,\% of the cluster radius, whereas this number amounts to the 40 most massive (23.5-3.2\,M$_{\odot}$) stars inside the same percentage of radius for Ber96.  This quantitative difference agrees with the suggestions on spatial distribution obtained from the density maps in Fig.\,\ref{Spatial}

The variation of Q with mass is plotted in the lower panel of Fig.\,\ref{lambdaQ}. There we see how Ber96 exhibits a radial distribution of the 17 more massive members (Q$\approx$ 0.96), whereas a high degree of substructure is observed in Ber94 (Q$\approx$  0.70) for the same number of stars. The limiting value Q$=$0.80 would separate radial from fractal spatial distributions (Cartwright \& Whitworth 2004). In this context we recall the important bias that spatial outliers could introduce on the estimation of Q. In our clusters this effect is however negligible. The calculation of Q values with rejection of the outmost 2 percent of stars (Parker \& Meyer 2012) leads to the same results as those shown in the lower panel of Fig.\,\ref{lambdaQ}.

The information on initial spatial distribution and early dynamical evolution derived from these parameters have been largely discussed in the literature. Ascenso et al. (2009) have suggested that crowding could lead to  undetection of relatively faint members, thereby biasing the results towards and overestimate of mass segregation of the most massive and bright members. Their conclusions mainly apply to the effects of important crowding in massive clusters, and can be considered to affect little to our clusters, which are one order of magnitude less massive and clearly less crowded (Fig.\,\ref{mapas}) than those simulated by these authors.

\begin{figure}
   \centering
   \includegraphics[width=10cm,height=12cm,angle=0]{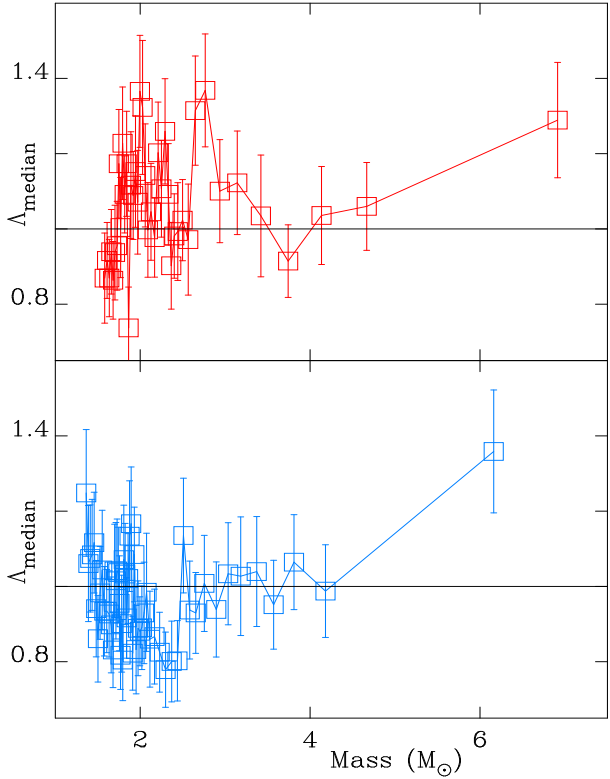}
      \caption{Values of the $\Lambda$ parameter, plotted in moving intervals of 25 stars, with steps of 5 stars, as a function of average mass in the interval. Red symbols (upper plot) correspond to Ber94, blue ones to Ber96. Horizontal lines mark the value $\Lambda$=1 (no mass segregation).}
         \label{L25}
   \end{figure}

However, $\Lambda$ parameter is very sensitive to the presence of outliers, in the sense that just one single star well separated from a high concentration of objects with similar masses could hide the mass-segregation effect, moving the $\Lambda$ parameter to values close to 1. Maschberger \& Clarke  (2011), and Parker et al. (2011) discuss how the methodology proposed by Allison et al. (2009) works well under some specific conditions but fails when dealing with samples including a few outliers. The solution proposed by these authors is to use the median instead the mean when determining the central value of the MST "edge length" distribution. We have analyzed both clusters with this procedure by calculating $\Lambda_{median}$ within a sequence of different mass ranges (instead of calculating a measure for all stars from the most massive down to the i-th most massive candidate members) and representing them versus the mass average in each interval.  We have repeated the the $\Lambda$ and Q calculations with 25 stars per bin, the number of stars per bin also selected for the study of the mass function. The results are shown in Figure\,\ref{L25}, where error bars represent the edge length dispersion divided by the edge length mean of the 25 random stars for each mass interval.  The presence of a weak but statistically significant mass segregation in both clusters for stars more massive than 6 M$_\odot$ solar masses can be inferred from these plots, a result which supports the conclusions from Fig.\,\ref{lambdaQ}.

\begin{figure}
   \centering
   \includegraphics[width=10cm,height=12cm,angle=0]{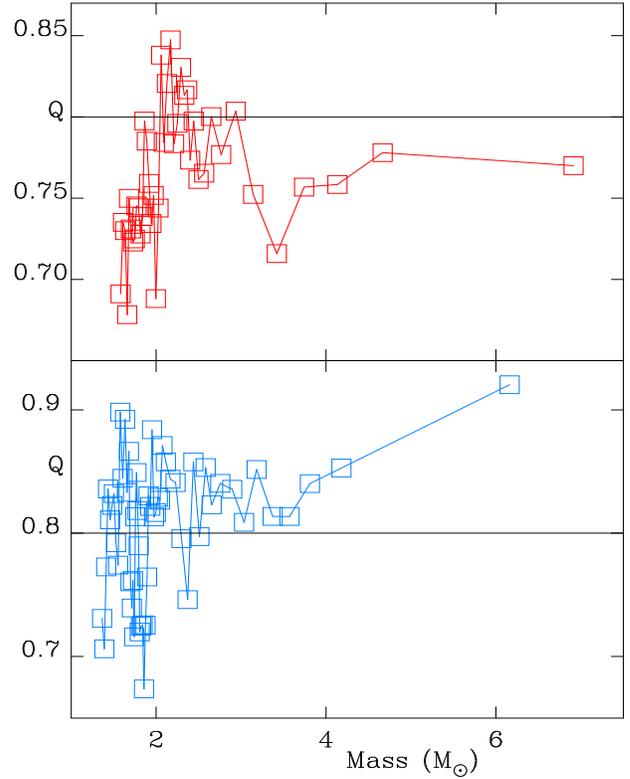}
      \caption{The same as Fig.\,\ref{L25}, for Q values. Horizontal lines mark the separation of spherically symmetric and substructured distribution (Q=0.8).}
         \label{Q25}
   \end{figure}

In the calculation of Q for different mass intervals, using this same procedure, we must consider the dependence that the results might have on the inclusion  of certain stars in any of two adjacent mass bins. The calculation in moving mass bins, with 25 stars per bin and steps of 5 stars, produces the results shown in Figure\,\ref{Q25}. They show that Ber96 is radially distributed, with Q values above 0.9 for the massive stars, whereas the opposite is observed for the most massive candidate members in Ber94, which exhibit a significant degree of structure. The results of this different approach support those suggested by the cummulative analysis. Stars with masses above 6 M$_\odot$ exhibit a comparatively higher concentration, although theis spatial distributions are different in both clusters.


The differences in geometry can be interpreted on the basis of the results from theoretical simulations (Parker \& Meyer 2012). The initial energy balance at the start of cluster formation can be characterized by the ratio Q$_{VIR}$ = T/$\left|\Omega\right|$|, where T and $\Omega$ are the total kinetic and potential energy of the stars, respectively. A fractal cluster initially supervirial (Q$_{VIR} > $ 0.5) would evolve expanding, but also keeping a high degree of substructure, visualized by the variation of Q, which remains well below the cut-off value of 0.8 up to 10 Myr (warm collapse). On the other hand, a similar cluster which starts with subvirial energetic balance (Q$_{VIR} < $ 0.5) would undergo a ``cold'' collapse, where the initial structure is rapidly erased, and massive stars are concentrated towards the cluster center.

Ber94 and Ber96 could be representatives of these two different dynamical evolutive paths. As it can be seen in Fig.\,\ref{lambdaQ}, Ber94 shows a distribution of the massive component with a high degree of structure (Q$\approx$  0.70) and is at the same time centrally concentrated. This could be explained by an early dynamical evolution under a supervirial regime. In the case of Ber96, its main characteristics suggest that the cluster has undergone a ``cold'' collapse instead. We remark that the present data do not allow us to make any inference about whether some primordial mass segregation was present in the clusters.

\subsection{Mass distribution}

A thorough analysis of the mass distribution in a cluster, and its completeness must take into account the influence of biases affecting the member selection and the mass assignments to them. Furthermore, even if the obtained mass distribution can be considered reasonably complete in the pertinent mass range, its interpretation in terms of a mass function established by the physics of the formation process needs consideration of several sources of bias of different qualities and importance (Kroupa et al. 2011, K11 in the following).

Among the biases described in K11, those biases that influence the actual calculation of a mass distribution from observed quantities are meant to be alleviated by our procedure. The restriction in age, colour excess and location in the Galaxy used to select our clusters are expected to minimize biases that arise from high and differential reddening, or crowding. A more persistent source of bias would indeed be the presence of unresolved multiple stars among the cluster members. On the other hand, the application of restrictions in the members selection, considering masses above an approximated completeness limit, and the use of equally populated bins in the calculation (Ma\'\i z-Apell\'aniz \& \'Ubeda 2005) are aimed at obtaining results as representative as possible of the actual mass distributions of the clusters. 

   \begin{figure}
   \centering
   \includegraphics[width=10cm,height=10cm,angle=0]{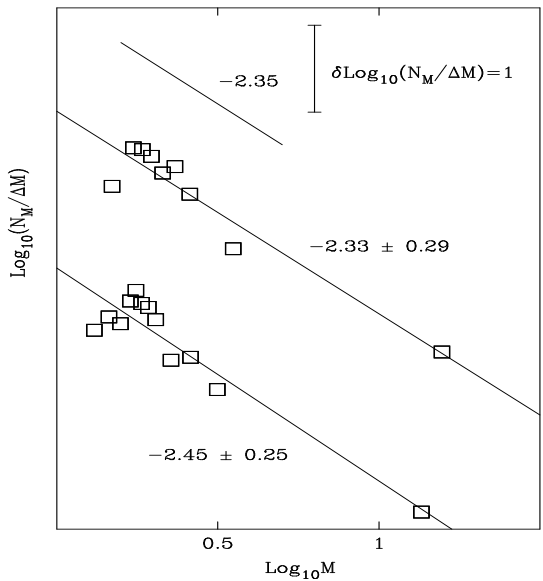}
      \caption{Mass distribution calculated for Ber94 (upper plot) and Ber96, using G02 and S00 models, and evenly populated bins, with 25 stars per bin. MS and postMS members are considered. Vertical error bars are similar to the squares size. The lines represent linear fits to the calculated distributions. Slopes with errors, and the vertical scale are shown. A segment with slope -2.35 is shown for comparison. An acceptable agreement inside the uncertainty is found with the slope of a Salpeter MF.}
         \label{MassDist}
   \end{figure}

With the calculated distance and colour excess, an approximate degree of completeness can be estimated. For each cluster, we estimate this level to be at the star whose absolute magnitude in the 5\,Myr PMS isochrone, added to the apparent distance modulus of the cluster (3.1$\times E(B-V)+DM$), equals the faintest magnitude of all observed stars with photometric errors in $V$ and in all four colour indices below 0.05 mag. The corresponding limiting mass amounts to 1.30\,M$_{\odot}$ and 1.35\,M$_{\odot}$ in Ber94 and Ber96, respectively. In practice, this includes all assigned member stars which also have detection in all bands (see Sect. 3.2 above). 

In Figure\,\ref{MassDist} we plot the obtained mass distributions for both clusters, with use of G02 and S00 models. The slopes and uncertainties are quoted. The slopes are to be compared with the one of the Salpeter function in linear form, {\sl dN$_M$/dM$\propto$M$^\alpha$} (Ma\'\i z-Apell\'aniz \& \'Ubeda 2005). A marginal overnumber of detected lower mass members could be present in Ber96, leading to a relatively steeper distribution. A general agreement inside the uncertainties is found between the slopes of the mass distributions from S00 models and the Salpeter (1955) mass function slope. 

The total masses of the proposed candidate members amount to 654$\pm$18 and 685$\pm$53 M$_\odot$ for Ber94 and Ber96, respectively. An estimate of the total mass, from our approximate limit, 1.3 M$_\odot$ down to 0.01 M$_\odot$, requires the assumption of a mass function. The Kroupa (2002) mass function leads to respective values for Ber94 and Ber96 of 2786 and 3455 M$_\odot$. Lower and upper limits, calculated with consideration of the uncertainties given for the slopes, amount to 1535, 6622 M$_\odot$ for Ber94, and 1934, 8151 M$_\odot$ for Ber96, which leave the total masses practically undetermined. We note however that the central total values calculated with the slopes proposed by Weidner, Kroupa \& Pflamm-Altenburg (2013) are sensibly smaller, with respective values of 1612 and 1897 M$_\odot$ for Ber94 and Ber96.

\section{Conclusions}

   \begin{enumerate}

      \item Photometric magnitudes and colours in the $UBVR_{C}I_{C}$ and $JHK_S$ systems are presented for 6150 stars in the fields of the Galactic open clusters Berkeley\,94 and Berkeley\,96. These constitute the first deep photometric study of these clusters. 

      \item Berkeley\,96 is found to be significantly younger than Berkeley\,94, with respective ages, LogAge(yr)=7.0$\pm$0.07 and LogAge(yr)=7.5$\pm$0.07. This finding is in agreement with the respective numbers of assigned  PMS members, and follows the trend between PMS member number and cluster age found previously (DAYII). 

      \item A population of sources with NIR excess in the $(H-K_{S})$ index is found in both fields. From their location in the $V$,$(V-I_{C})$ CM diagram, for the fraction that is optically visible, and mainly from the spatial distribution that they exhibit, we suggest that some of them could represent the lowest mass end of the cluster member sequence. The suggestion is particularly strong in the case of the younger Berkeley\,96, where the number of NIR-excess sources found is correspondingly higher. This would imply the presence in the cluster of member stars of age around 10 Myr, covering a mass range of 2 orders of magnitude, from around 27 to 0.2\,M$_\odot$.

      \item The spatial distributions of candidate members in various mass ranges suggest different initial conditions and formation processes in these clusters. A quantitative analysis of the features observed in the obtained spatial distribution, is performed according to published numerical simulations of cluster evolution. The results suggest the presence of mass segregation in both clusters. On the other hand we could be witnessing results of the so-called ``warm'' and ``cold'' collapse models, which would be represented in the respective formation and dynamical evolution of Berkeley\,94 and Berkeley\,96.

      \item Mass distributions are calculated for both clusters with mass values from G02 and S00 models. Conservative restrictions for the candidate members are applied to avoid biases at the low mass end. Bins with evenly distributed star number are used. The slopes of the calculated mass distributions show an acceptable agreement within the uncertainties with the Salpeter MF slope. 

   \end{enumerate}

\section*{Acknowledgments}

The NOT research students S\o ren Frimann, Julie Lykke, Ricky Nilsson, Anders Thygesen and Paul Wilson are warmly acknowledged for performing part of the ALFOSC observations. This work has been supported by the Spanish Ministerio de Educaci\'on y Ciencia, through grant AYA2010-17631, and by the Consejer\'\i a~ de Educaci\'on y Ciencia de la Junta de Andaluc\'\i a, through TIC101 and P08-TIC-4075. The data presented here were obtained in part with ALFOSC, which is provided by the Instituto de Astrofisica de Andalucia (IAA) under a joint agreement with the University of Copenhagen and NOTSA. We made use of the NASA ADS Abstract Service and of the WEBDA data base, developed by Jean-Claude Mermilliod at the Laboratory of Astrophysics of the EPFL (Switzerland), and further developed and maintained by Ernst Paunzen at the Institute of Astronomy of the University of Vienna (Austria). This publication makes use of data products from the Two Micron All Sky Survey, which is a joint project of the University of Massachusetts and the Infrared Processing and Analysis Center/California Institute of Technology, funded by the National Aeronautics and Space Administration and the National Science Foundation. This publication makes use of data products from the Wide-field Infrared Survey Explorer, which is a joint project of the University of California, Los Angeles, and the Jet Propulsion Laboratory/California Institute of Technology, funded by the National Aeronautics and Space Administration. We acknowledge the use of NASA's SkyView facility, located at NASA Goddard Space Flight Center.

\end{document}